\documentstyle[twoside,fleqn,espcrc2,epsfig]{article}
\title{Cosmic ray composition and energy spectrum above 1 TeV: direct and
EAS measurements}

\author{A.Castellina\address{Consiglio Nazionale delle Ricerche, Istituto
        di Cosmogeofisica, cs.Fiume 4, 10133 Torino, Italy}%
        \thanks{E-mail: castellina@to.infn.it.}}

\begin{document}

\begin{abstract}
The most recent experimental results on the cosmic ray composition and energy
spectrum above 1 TeV are reviewed and discussed. 
All data agree on the presence of the so-called ``knee'' at an energy
$E_{k} \simeq 3 \ 10^{15} eV$; the knee is seen in all the 
components of the Extensive Air Showers. These results support the hypothesis
of an astrophysical origin of the knee, while no new features  in the
hadronic interactions at high energies are envisaged.\\
The cosmic ray composition below and above the knee region is still an open
question. According to most experiments, the knee seems to be due to the
light component of the primary beam, with a composition getting heavier
above the knee. However, results contradicting this conclusion have to be
considered and understood.
\end{abstract}

\maketitle

\section{Introduction}

The energy spectrum of cosmic rays (CR) spans a very wide energy
range, with particle fluxes steeply falling more than 30 orders of
magnitude. Above the solar modulation region, the spectrum can be well
described by a power law, which steepens around $3 \ 10^{15} eV$, a feature
called the ``knee'', discovered in 1958 \cite{kri}; it softens again at
$\simeq 10^{19} eV$, the ``ankle''.\\ 
Explaining the knee feature would shed light on the CR origin and
acceleration mechanism, depending on whether it is a signature of a change
in the hadronic interactions at such energies or it reflects a feature of
the cosmic ray spectrum, thus concerning mainly astrophysics.\\
Several arguments involving energetics, composition and secondary $\gamma$
ray production suggest that cosmic rays at least up to the knee region are
confined in the Galaxy. \\
The most popular theory is that of diffusive shock acceleration in
Supernova remnants (SNR), that is particle acceleration by SNRs expanding
supersonically in the surrounding medium. \\
Supernova explosions can easily account for the energy stored in galactic
cosmic rays; the spectrum emerging from the SNR is of the type $E^{-2.1}$
up to a maximum energy near $10^{14}$ eV times the nuclear charge, after
which it drops very rapidly.
Folding the production spectrum with the effect of diffusion through the
Galaxy, and taking a trapping time varying as $E^{-0.6}$ (as found from the
proportion of secondary to primary nuclei arriving to Earth), the resulting
flux of CR in the Galaxy would be $\propto E^{-2.7}$, in close
agreement with expectations. The maximum achievable energy is close to the
knee one.\\
A direct evidence that the nucleonic component of CR is indeed produced in
SNRs could be obtained by the observation of $\gamma$ rays: the accelerated
cosmic rays can in fact interact with the local interstellar matter, in
this way producing $\gamma$ rays by either hadronic or leptonic production.\\
Various experimental groups reported on TeV $\gamma$ emission from
supernova remnants, like SN1006, RJX1713.7-3946, Cassiopea-A (see
e.g. \cite{weekes} and references therein). Unfortunately in all cases the
emission can be attributed to electron progenitors and no
positive evidence for hadroproduction of TeV $\gamma$'s has been found yet.\\
Further information can be obtained from the study of the distribution
of CR arrival directions. A recent compilation of the anisotropy
measurements can be found in \cite{anis}; 
while the amplitude and phase of anisotropy data below $\simeq 2 \ 10^{14}
eV$ are consistent and statistically accurate \cite{et-ani}, the
experimental results at higher energies are not compatible with
expectations. This could mean that the diffusion model cannot be simply
extrapolated to these energies \cite{hillas}.\\
Various models have been put forward trying to identify the sites and
mechanisms of injection of cosmic rays at higher energies, at and above the
knee. If the bend in energy spectrum is related to the maximum
achievable energy in the accelerator, then CR at higher energies could be
powered by a reacceleration by interstellar turbulence \cite{Seoptu}, or
they could be produced by Supernovae exploding in denser media (their stellar
wind cavity) (\cite{Bie98} and references therein).
On the other hand, the knee could be attributed to propagation effects.
In both cases, one would expect multiple bends due to the different
elements bending at fixed rigidity; the composition would become heavier
above the knee.
An extra-galactic origin for CR above the knee has also been proposed
\cite{prot}, where the accelerator sites are found in Active Galactic
Nuclei and the resulting composition is getting lighter above the knee.\\
A completely different point of view assigns the knee to a new dramatic
process of hadronic interaction which takes over around the knee
energy. However, even if it is true that in the $\Delta E$ of interest we
have no direct information about the hadronic interaction cross section for
the secondary production relevant to the interpretation of measurements, no
experimental data as far show a need for a different interaction mechanism.\\
From the experimental point of view, what is most important in order to
test the models is to measure the cosmic ray composition and energy spectrum
near the energy limit of the shock models; moreover, measurements of
anisotropy and secondary to primary ratio at higher energy are of utmost
importance.

\section{Direct measurements}

\begin{figure}[b]
 \begin{center}
 \mbox{\epsfig{figure=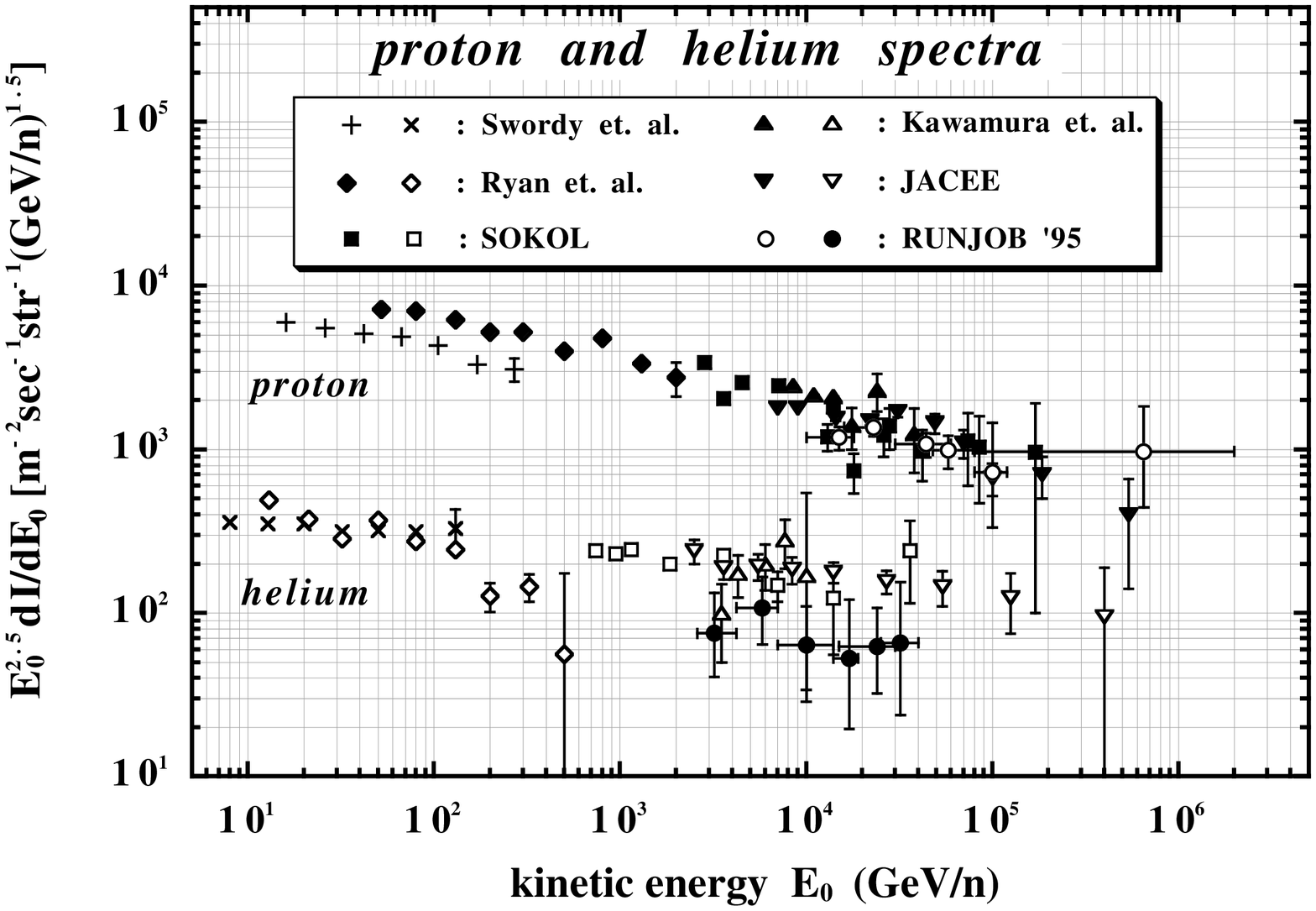,width=7cm,height=6cm}}
 \end{center}
\caption{\em{Differential energy spectra for proton and helium \protect\cite{shiba}.}}
\label{fi:spephe} 
\end{figure}

\begin{figure}[!htb]
 \begin{center}
   \mbox{\epsfig{figure=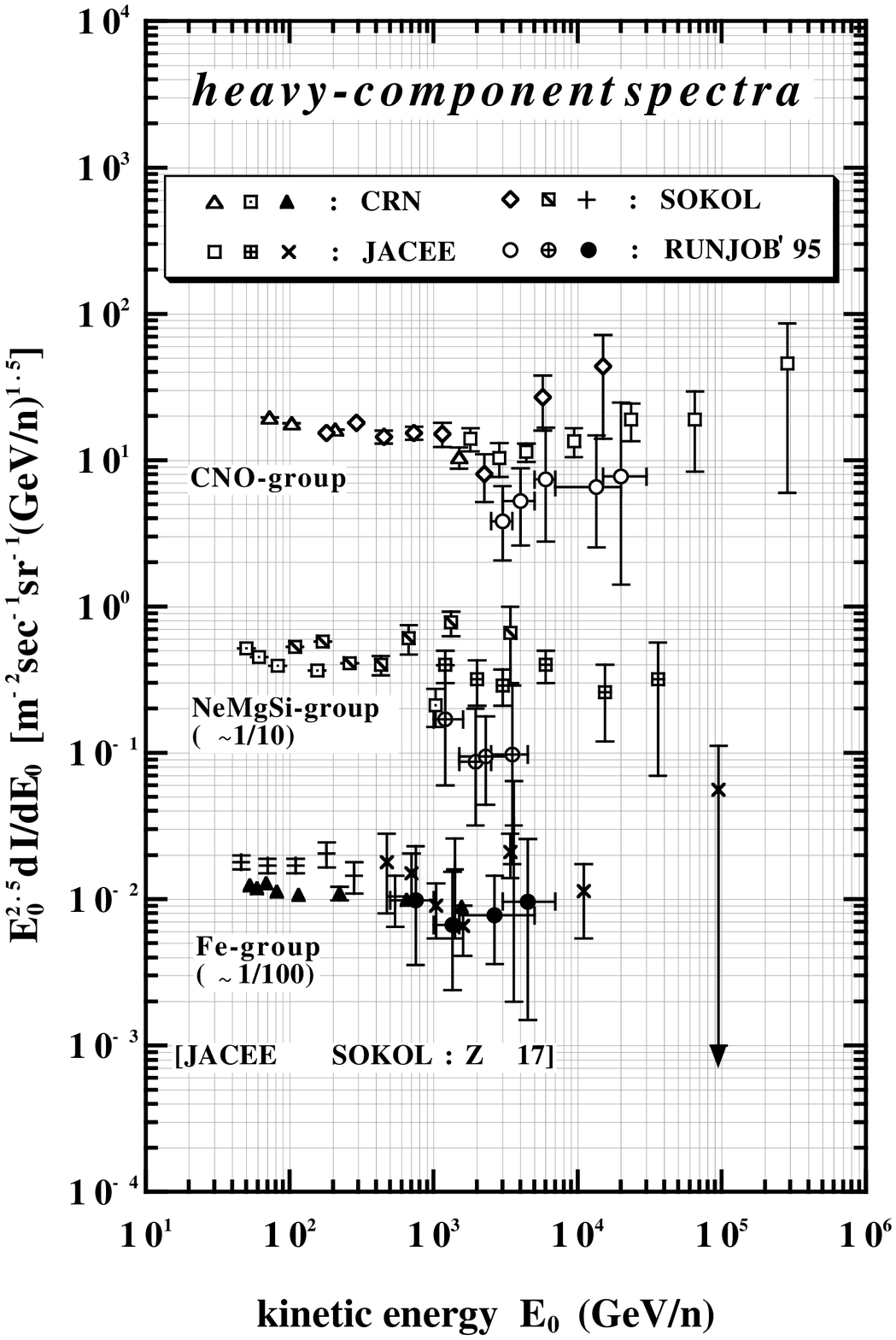,width=7cm,height=6cm}}
 \end{center} 
 \caption{\em{Differential energy spectra for the CNO, NeMgSi and Iron groups
\protect\cite{shiba}.}}
\label{fi:spefer}
\end{figure}

\begin{figure}[!htb]
 \begin{center}
 \mbox{\epsfig{figure=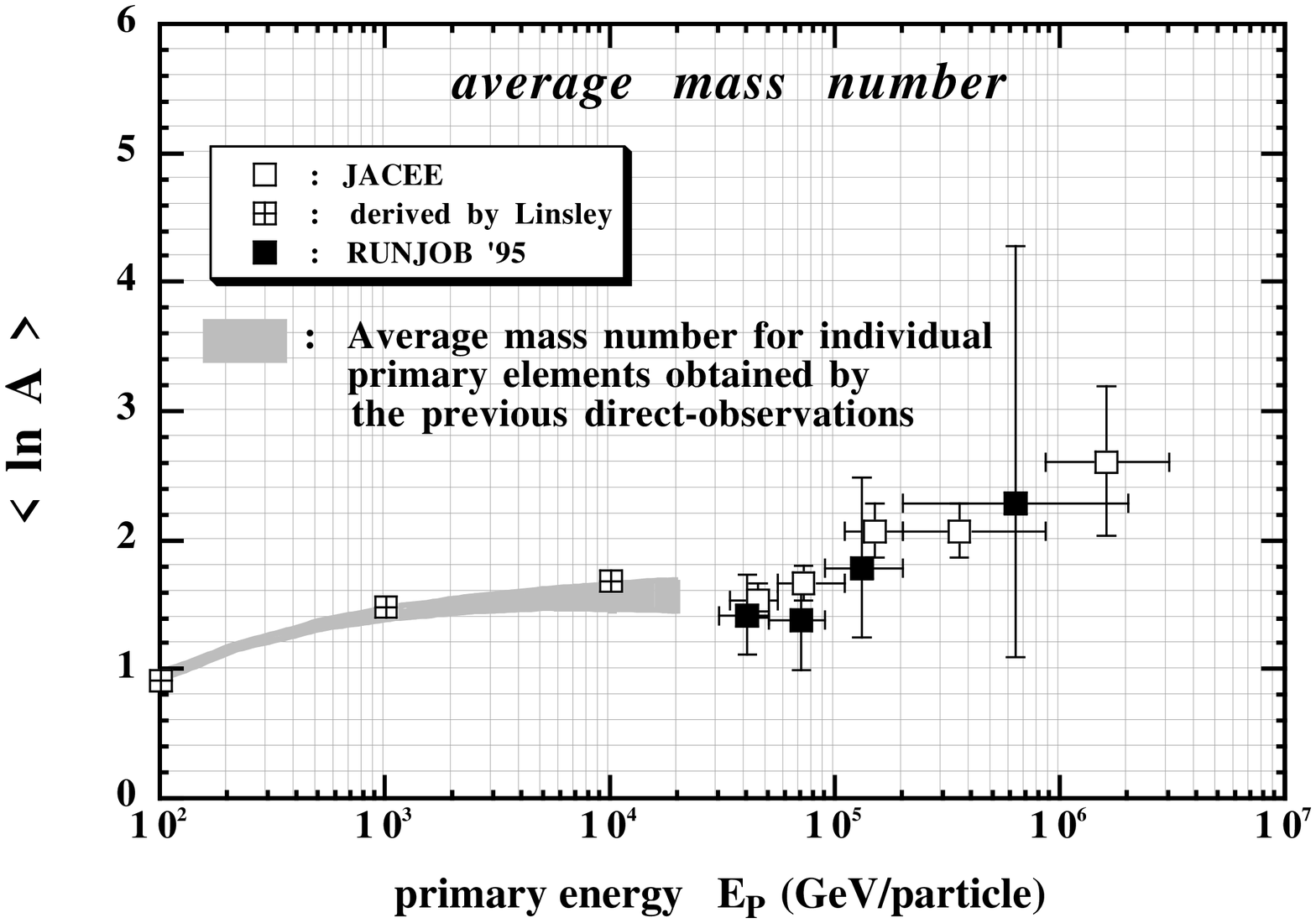,
width=7cm,height=6cm}}
 \end{center} 
\caption{\em {Average primary mass vs primary energy from direct
measurements \protect\cite{shiba}.}}
\label{fi:dirlna} 
\end{figure}

Direct measurements of the relative abundances of the
cosmic ray nuclei and their distribution in energy are possible only at
relatively low energy: they in fact require installation of instrumentation
on balloons or space shuttles flying outside the atmosphere at very high
altitude. 
The most recent results still come from two balloon experiments, JACEE and
RUNJOB, as summarised in \cite{wats,shiba}.\\
The proton and Helium spectra have been measured by JACEE up to about 800 TeV
\cite{jacee}; no break was found in the proton one, but above 80-90 TeV the
experiment does not have enough statistics to either assess or reject its
presence \cite{jac99}.\\
JACEE claim for a flatter He spectrum as compared to the proton one is
in agreement with SOKOL result \cite{sokol}, but this is not confirmed by the
results by RUNJOB \cite{runj99}. The JACEE group reported values are
$\gamma_{p} = (-2.80 \pm 0.04)$ and $\gamma_{He} = (-2.68 \pm 0.06)$, while
RUNJOB slopes are both $\simeq -2.80$ with an uncertainty between $10$ and 
$20 \%$ for both protons and Helium nuclei.
It should however be mentioned that the significance of the difference
between the slopes for p and He is at the level of only $2 \sigma$; on the
other hand, the results from RUNJOB are based only on the 1995 and 1996
data ($\simeq 30 \%$ protons and $\simeq 13 \%$ He respect to JACEE data).
The experimental results on the p and He slopes are of particular importance
as regards the models of non linear acceleration of cosmic rays, where the
injection rate is an increasing function of the primary particle rigidity 
\cite{bere}.\\
The single component spectra are shown in Fig.\ref{fi:spephe} and in
Fig.\ref{fi:spefer}. 
The experimental results agree for what regards the iron spectrum,
while RUNJOB gives a factor of 2 lower spectra for the C-N-O and Ne-Si
groups. Data are all consistent with an increase of the mean logarithm
of the average primary mass $<ln \ A>$ with energy, as shown in 
Fig.\ref{fi:dirlna}.


It is clear that more statistics is needed above $100 \ TeV$, and various new
projects are under development.
ACCESS \cite{access} is estimated to be launched on the International Space
Station in 2006.
Its primary goal will be the measure of energy spectrum and composition up
to $10^{15} eV$, thus testing Supernova shock acceleration models; the
charge range at high energy will be $1 \leq Z \leq 28$. Three different
detectors are being built for this purpose: a charge identification module,
to measures the abundances of all individual elements, a transition
radiation detector, to identify and measure the energy of particles with $Z
\geq 2$ up to $\simeq 100 \ TeV/nucleon$, a calorimeter to measure the
particle energies and to identify electrons.\\
The ATIC \cite{atic} project, in its initial design for long duration
balloon flights, is devoted to study the energy spectrum of
Galactic proton and helium up to $10^{14} eV$, in order give information
about the proton/helium ratio, the possible difference in their spectral
slopes, the existence of a bend in the proton spectrum.\\
CREAM \cite{cream} plans to explore spectrum and composition up to 
$\simeq 10^{15} eV$, exploiting ultra long duration balloon flights
($\simeq 100$ days). With an exposure of $\simeq 300 \ m^{2}sr \ days$,
this instrument will collect $\simeq 500$ proton and helium nuclei above
$10^{14} eV$, reaching $\simeq 30 \%$ statistical accuracy above $10^{15}
eV$.\\
The new Ionization Neutron Calorimeter INCA \cite{inca} proposes to study
the range
0.1-10 PeV using the well known techniques of ionization and neutron
monitor to measure energy and a silicon particle charge detector 
to determine the charge and coordinates of the primaries.\\
Combining different detectors, the new projects will have a very powerful
tool to overcome the individual technical limitations.

\section{The energy spectrum}

\begin{figure}[htb]
 \begin{center}
 \mbox{\epsfig{figure=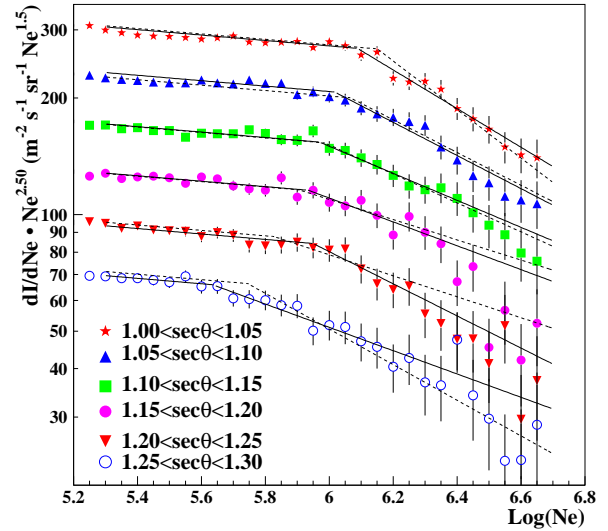,width=8cm,height=8cm}}
 \end{center} 
\vskip -0.5cm 
\caption{\em {Differential electron size spectrum at different zenith
angles, that is different atmospheric depths, as measured by EAS-TOP. The
solid lines show the results of the fitting procedure \protect\cite{et-spe}; the
fits represented by dashed lines were obtained requiring a constant
integral flux above the knee.}}
\label{fi:etesiz} 
\end{figure}

The experimental observables which are measured in order to extract
information about the energy spectrum are the charged components of showers as
measured by ground based detectors with scintillator counters, muon and
hadron detectors, or the $\breve{C}$erenkov light produced by shower
particles as they propagate through the atmosphere.\\
The interpretation of these ground level observations in
terms of primary particle characteristics is far from straightforward,
being strongly dependent on models simulating the production and
propagation of particles through the
atmosphere. Models in turn depend on extrapolations applied to the data on high
energy particle interactions studied at accelerators; the energy
region of interest is in fact much higher than that studied at
accelerators, the explored kinematic region is the forward one, the
collisions among nuclei make the influence of nuclear effects not
negligible.\\
The most recent results of EAS experiments concerning the primary energy
spectrum of cosmic rays are described in the following.\\
The electron size spectra as measured by the EAS-TOP
experiment \cite{et-gen} are shown in Fig.\ref{fi:etesiz}. The knee is
clearly visible, and the size corresponding to the knee shifts towards 
lower values at increasing atmospheric depth, as expected
for a knee at given primary energy \cite{et-spe}.

The shower size at the knee decreases with increasing atmospheric depth, 
with an attenuation length $\Lambda_{k}=(222 \pm 3) \ g \ cm^{-2}$, in very
good agreement with that found for the shower absorption in atmosphere; the
integral intensities $I_{k}(\geq E_{k})$ are constant within $20 \%$.
The knee in electron size is quite sharp, showing that the change in slope
occurs in a limited range of $N_e$ ($\Delta N_{e}/N_{e} \leq 25 \%$).

\begin{figure}[htb]
 \begin{center}
 \mbox{\epsfig{figure=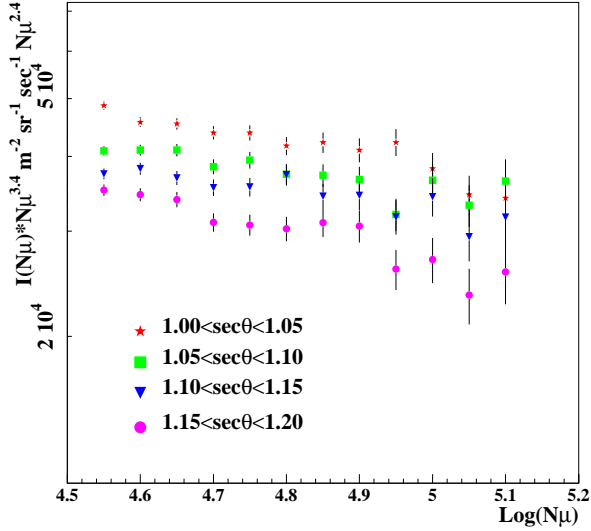,width=8cm,height=8cm}}
 \end{center} 
\vskip -1.0cm 
\caption{\em {Differential muon size spectrum at 4 different atmospheric
depths as measured by EAS-TOP \protect\cite{mumin}.}}
\label{fi:etmusiz} 
\end{figure}

\begin{figure}[!htb]
 \begin{center}
 \mbox{\epsfig{figure=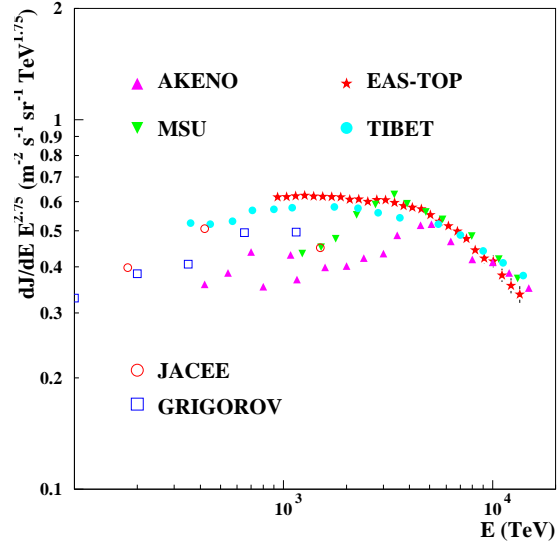,width=8cm,height=8cm}}
 \end{center} 
\vskip -1.0cm 
\caption{\em {Primary energy spectrum as obtained by EAS-TOP compared with
other experimental results \protect\cite{et-spe}.}}
\label{fi:etspe} 
\end{figure}

The muon size spectra have been measured in 4 different zenith angle
intervals, as one can see in Fig.\ref{fi:etmusiz}, where the change of
slope is visible at all atmospheric depths despite the large statistical
fluctuations. The knee, around $N_{\mu}^{k} \simeq 10^{4.65}$, is
independent on the number of detected muons, being in
fact visible at any core distance \cite{mumin}.
The integral fluxes in electron and muon size are compatible at all
atmospheric depths, as expected for a feature occurring at fixed primary
energy, also confirming the consistency of the whole procedure.\\
A further interesting result comes from the relation between the electron
and muon size slopes, which can be written as $N_{\mu} \propto N_{e}^{\alpha}$ 
with $\alpha \simeq 0.75$ in all angular bins: no sudden change in the
secondary production when going through the knee region is seen, thus
showing that, at least  from this point of view, no new hadronic effects is
needed to explain the knee \cite{etalf}.\\
A simulation of the shower production and development in atmosphere using
the CORSIKA code \cite{corsika} with the HDPM interaction model allows to
find the relation between shower size and primary spectrum 
$N_{e}(E_{0},A)=\alpha(A_{eff}) E_{0}^{\beta(A_{eff})}$
The effective mass $A_{eff}$ is calculated from the extrapolation of
the single nuclear spectra measured at low energies by direct measurements;
above the knee, a rigidity dependent cutoff is used. \\
The final result is shown in Fig.\ref{fi:etspe}; the agreement with direct
measurements at low energies and with other air shower experimental
results at the highest ones is quite good. The systematic uncertainties
in the energy spectrum are due to the primary composition and interaction model
chosen in the calculation: the maximum difference in the determination of
$N_e$ between different interaction models and HDPM is $\simeq 10 \%$; the
all-particle flux obtained with ``heavy'' or ``light'' limit compositions
differs by $\simeq 10 \%$ from the above calculated one.

The energy spectrum is determined by CASA-MIA using the muon $N_{\mu}$ and
electron $N_{e}^{*}$ size measurements \cite{casa-spe}. The $N_{e}^{*}$
indicates in this case the sum of $e^{+},e^{-},\gamma$ at the ground.
The sizes combination $F=log_{10}(N_{e}^{*}+\psi N_{\mu})$ was found to be
log-linear in $E_0$ and, what is most important, independent on the primary
mass. 

\begin{figure}[htb]
 \begin{center}
 \mbox{\epsfig{figure=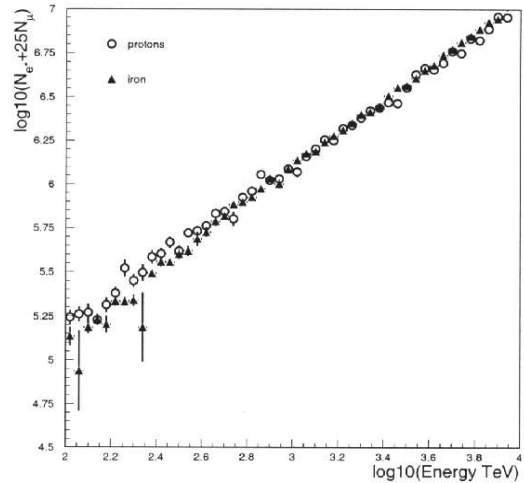,width=7cm,height=6.5cm}}
 \end{center} 
\vskip -0.5cm 
\caption{\em {$F=log_{10}(N_{e}^{*}+\psi N_{\mu})$ as a function of energy
for simulated proton (open circles) and iron (black triangles) primaries
\protect\cite{casa-spe}.}}
\label{fi:cacomb} 
\end{figure}

\begin{figure}[htb]
 \begin{center}
 \mbox{\epsfig{figure=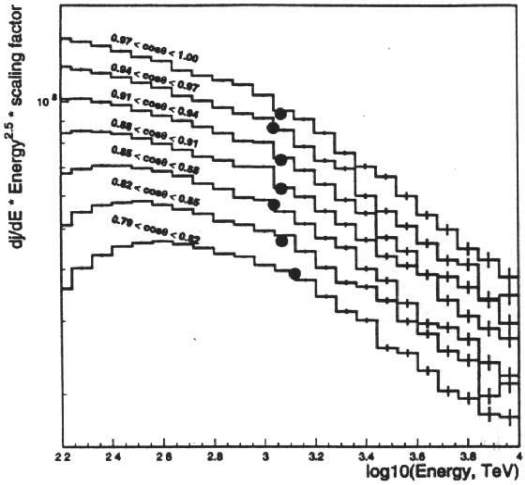,width=7cm,height=6.5cm}}
 \end{center} 
\vskip -1.0cm 
\caption{\em {Primary energy spectrum as found by CASA-MIA \protect\cite{casa-spe}.}}
\label{fi:casaspe} 
\end{figure}

In Fig.\ref{fi:cacomb} this relation is shown as found from a
simulation of primary protons or iron; the model used was the QGSJET
one.
The systematic differences in energy assignment for different primary mass
A are less than $5 \%$. The average absolute energy reconstruction errors go
from $\simeq 25 \%$ at $10^{14} eV$ to $\simeq 16 \%$ at $\geq 10^{15} eV$.\\
The parameter $\psi$, which defines the relative weight of muons and
electrons in the showers, is strongly dependent on the model used for hadronic
interactions, but the change in energy assignment due to this effect is
claimed to be $\leq 10 \%$. 
The mass insensitivity allows to determine the energy free of
systematic effects (on the contrary, for example, if in some region the
energy spectrum changes, $N_e$ vs E also changes).\\
The energy spectrum thus derived is shown in Fig.\ref{fi:casaspe}; 
the knee is located at the same primary energy for any atmospheric depths,
as expected. 

\begin{figure}[htb]
 \begin{center}
 \mbox{\epsfig{figure=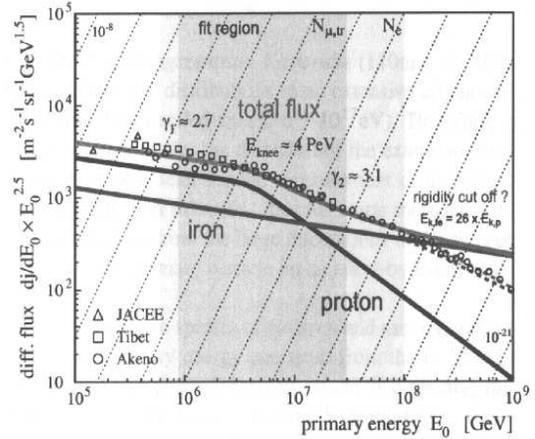,width=7cm,height=6cm}}
 \end{center} 
\vskip -1.0cm
\caption{\em {Primary energy spectrum by KASCADE from electron
and muon size spectra \protect\cite{kas-gl}.}}
\label{fi:kamusp} 
\end{figure}

KASCADE \cite{ka-gen} measures all the three charged components of
extensive air showers: electrons, muons and hadrons.\\
The knee is clearly visible in all components; for electrons and muons,
the higher statistics allows to study the size at different atmospheric
depths, thus finding that the size at the knee decreases at increasing
atmospheric depth \cite{ka-knee}. \\
The primary energy spectrum can be extracted from the measured size spectra 
depending on the knowledge of the mass composition as obtained from the
observables under investigation and on the relation between size and energy
resulting from simulation.\\
The energy spectrum shown in Fig.\ref{fi:kamusp} was found by a combined
$\chi^2$ minimisation to fit both the $N_{e}$ and the $N_{\mu}$ truncated
muon size spectra simultaneously (the truncated muon number is that found
by fitting the muon lateral distribution within a limited range  of 40-200
m) \cite{kas-gl}. The evaluated size spectra are in fact the
convolution of the energy spectrum and a kernel function describing the
probability of a given primary to produce a shower with a certain size and
which includes the parametrisations of shower fluctuations for both proton
and iron primaries according to Monte Carlo. The knee is found at about 4 PeV.

\begin{figure}[htb]
 \begin{center}
 \mbox{\epsfig{figure=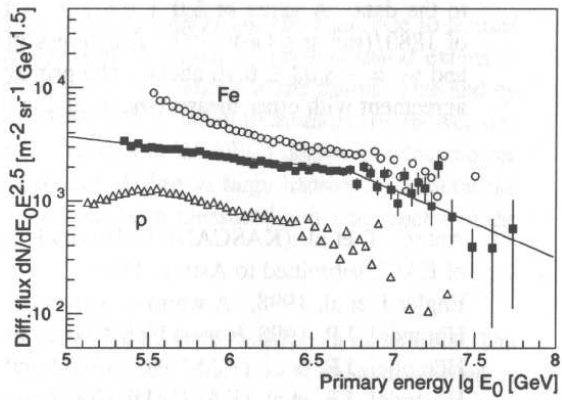,width=1.\linewidth}}
 \end{center} 
\vskip -1.0cm
\caption{\em {Primary energy spectrum as found using KASCADE hadrons.
Empty circles and triangles are the spectra from simulation of pure
iron or proton primaries respectively \protect\cite{ka-all}.}}
\label{fi:kahasp}
\end{figure}

\begin{figure}[htb]
 \begin{center}
 \mbox{\epsfig{figure=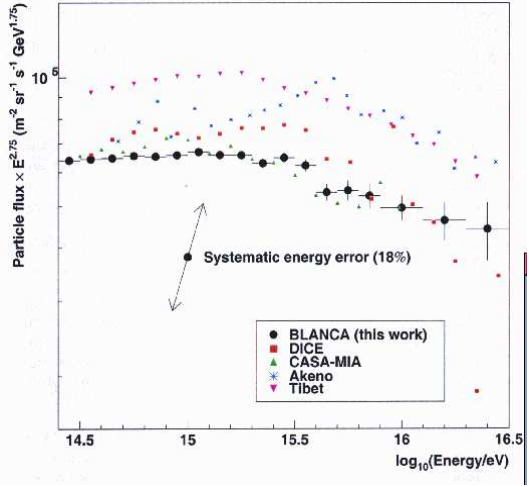,width=7cm,height=6.5cm}}
 \end{center} 
\vskip -1.0cm 
\caption{\em {Primary energy spectrum from BLANCA \protect\cite{bla-gen}.}}
\label{fi:blaspe}
\end{figure}

\begin{figure}[htb]
 \begin{center}
 \mbox{\epsfig{figure=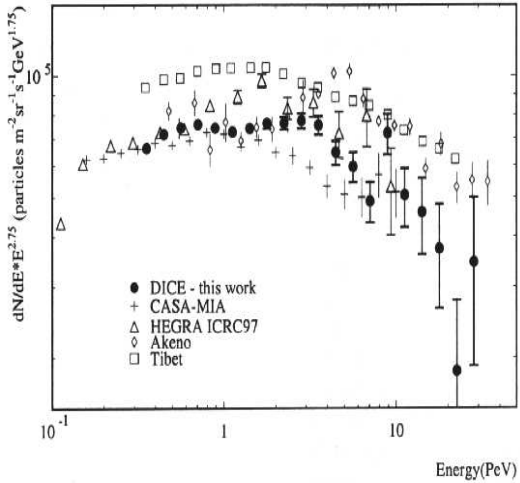,width=7.cm,height=6.5cm}}
 \end{center} 
\vskip -1.0cm 
\caption{\em {Primary energy spectrum from DICE events in coincidence with
CASA-MIA(\protect\cite{dice-gen}).}}
\label{fi:dicespe}
\end{figure}

In Fig.\ref{fi:kahasp}, the energy spectrum as derived from hadronic data
is also shown. The knee is again very clear, and the expectations from pure
beams of protons or iron primaries are shown for comparison.
Furthermore, KASCADE data show evidence of the knee also in the energy sum
of hadrons in the calorimeter \cite{ka-all}.\\
From muon density measurements in the multi-wire proportional chambers
below the central detector \cite{ka-mu}, a subdivision of the data in
``light'' and ``heavy'' samples (according to the parameter 
$Log \ N_{\mu}/Log \ N_{e}$ as described in Sect.4) shows that the knee is
strongly dominated by the light component, within a $30 \%$ uncertainty
due to Monte Carlo statistics.\\
The broader lateral distribution of the $\breve{C}$erenkov light, due to
the smaller absorption of photons in atmosphere, and the high photon
number density, that means a better signal-to-noise ratio even for smaller
arrays, are the main advantages in using $\breve{C}$erenkov detectors as
compared to charge particle counting arrays. 
The most recent results from apparata based on the detection of
$\breve{C}$erenkov light from Extensive Air Showers come from BLANCA
\cite{bla-gen} and DICE \cite{dice-gen}, both operating at the same site
and sharing some equipment with CASA.\\
The first one consists of 144 angle-integrating $\breve{C}$erenkov light
detectors located in the CASA scintillator array, which provides the
trigger and gives core position and shower direction.
The $\breve{C}$erenkov lateral distribution function is measured and fitted
through the expression $C(r) = C_{120} e^{-sr}$ in the inner part of
the distribution ($r \leq 120 \ m$). 
The intensity at a critical radial distance of 120 m, entirely determined by
density and scale height of the atmosphere, is proportional to the primary
energy and the dependence on the primary mass is fully included in the
slope $s$ of the distribution, which is in fact a function of the depth of
maximum development $X_{max}$.\\
DICE consists of 2 imaging telescopes of 2m diameter. What is measured is
the $\breve{C}$erenkov light size $N_{\gamma}$, by summing the total amount
of light at each phototube and the
depth of maximum development of the shower $X_{max}$ by fitting the shape
of the light image in each telescope, with a procedure that is essentially
geometrical and not depending on simulations, except for calculations to
determine the angular distribution of light around the axis. The core
position and shower direction are given by the CASA scintillator array.\\
The primary energy is estimated through a fit including geometry,
$N_{\gamma}$ and $X_{max}$ and takes therefore into account the
dependence of the lateral distribution and intensity of
the $\breve{C}$erenkov light, at fixed primary energy, on the primary mass.\\
The resulting energy spectra from BLANCA and DICE are shown in
Fig.\ref{fi:blaspe} (QGSJET model) and Fig.\ref{fi:dicespe} in comparison
with other experimental results.
The knee feature is evident in the BLANCA energy spectrum at 
$\simeq 3 \ PeV$; a $10 \%$ shift in the energy scale, which is however
less than the instrumental uncertainty, is found by changing the
interaction model chosen to interpret the data.
The absolute calibration error results in a $\simeq 18 \%$
systematic error on the energy assignment. 
According to DICE data, the knee is found around 3 PeV; 
the systematic uncertainty in the absolute flux is $\simeq 30 \%$, due to
the intrinsic error in the energy scale of $\simeq 15 \%$.


\begin{figure}[!htb]
 \begin{center}
 \mbox{\epsfig{figure=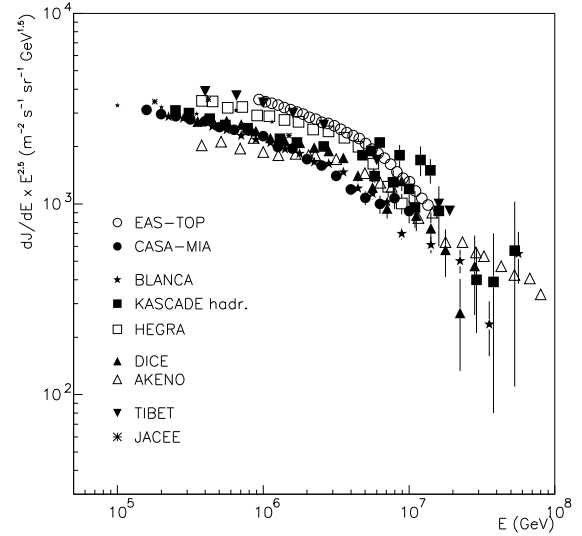,width=8cm,height=8cm}}
 \end{center} 
\vskip -1.0cm 
\caption{\em {All-particle primary energy spectrum from the various
experiments as described in the text.}}
\label{fi:spetutti}
\end{figure}

The $\breve{C}$erenkov lateral distribution has also been measured by HEGRA
with its AIROBICC detectors \cite{heg_1}. Again, the
scintillator array provides the core position and arrival direction of the
showers.
The total primary energy can be reconstructed by the electromagnetic one,
if one assumes a primary composition; however, a determination of the
primary energy in a mass independent way can be obtained following the
approach of Lindner \cite{lind}, at the expenses of the energy resolution,
which is worse than that found by the mass-dependent method and of a
stronger dependence of the result on the fluctuations of $X_{max}$.
The knee is found at about 3 PeV.\\
In Fig.\ref{fi:spetutti}, the all-particle primary energy spectrum as
obtained with the various experiments here described is shown. 
A summary of the up-to-date situation is given in Table
\ref{ta:tabslop}. The differences among the quoted knee energies are mainly
due to the assumed composition, which in turn depends on the
observables used as will be discussed below, but the existence of the knee
is clearly established between 2 and 5 PeV.

\begin{table*}[!htb]
\setlength{\tabcolsep}{1.5pc}
\caption{\em{Slopes of the energy spectrum below ($\gamma_1$) and above
($\gamma_2$) the knee. $E_k$ is the energy at which the knee is seen
(PeV).}}
\label{ta:tabslop}
\begin{tabular*}{\textwidth}{@{}l@{\extracolsep{\fill}}rrrr}
\hline
{\bf Experiment } & {\bf $\gamma_1$ } & {\bf $\gamma_2$ } & {\bf $E_{k}$ } \\ 
 & & & (PeV)  \\ \hline
\hline
 EAS-TOP \cite{et-spe} & $-2.76 \pm 0.03$ & $-3.19 \pm 0.06$ & $2.7 - 4.9$ \\
 KASCADE \cite{ka-em} & $-2.70 \pm 0.05$ & $-3.10 \pm 0.07$ & $4.0 - 5.0$ \\
 KASCADE \cite{ka-all} & $-2.66 \pm 0.12$ & $-3.03 \pm 0.16$ & $5.0 \pm 0.5$ \\
 CASA  \cite{casa-spe} & $-2.66 \pm 0.02$ & $-3.00 \pm 0.05$ & smooth \\
 AKENO \cite{akeno} & $-2.62 \pm 0.12$ & $-3.02 \pm 0.05$ & $\simeq 4.7$ \\
 TIBET \cite{tibet} & $-2.60 \pm 0.04$ & $-3.00 \pm 0.05$ & smooth \\
 TUNKA \cite{tunka} & $-2.60 \pm 0.02$ & $-3.00 \pm 0.06$ & $\simeq 4.0$ \\
 BLANCA \cite{bla-gen} & $-2.72 \pm 0.02$ & $-2.95 \pm 0.02$ & $2.0^{+0.4}_{-0.2}$ \\
 DICE \cite{kiswo} & $\simeq -2.7$ & $\simeq -3.0$ & $\simeq 3.0$ \\
 HEGRA \cite{heg_1} & $-2.67 \pm 0.03$ & $-3.33^{+0.33}_{-0.41}$ & $3.4^{+1.3}_{-0.7}$ \\ \hline
\end{tabular*}
\end{table*}

\section{Composition}

\begin{figure}[htb]
 \begin{center}
 \mbox{\epsfig{figure=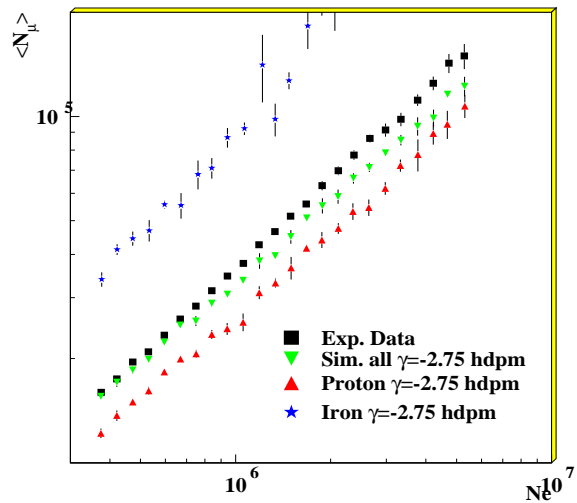,width=8cm,height=7.5cm}}
 \end{center} 
\vskip -1.0cm 
\caption{\em {$<N_{\mu}>$ vs $N_e$ from EAS-TOP experimental data (full
squares) as compared to a simulation with mixed composition and all
components with the same slope (downward triangles), pure proton (upward
triangles) and pure iron (stars) primaries \protect\cite{et-comp}.}}
\label{fi:etcomp}
\end{figure}

The cosmic ray primary composition measurements around the knee are crucial
for the understanding of the mechanisms of acceleration and the source
problem. \\
The experimental observables we are dealing with are: 
a) the $\breve{C}$erenkov lateral distribution or the image of the
$\breve{C}$erenkov light emitted by the shower in atmosphere; 
they allow to determine the depth of shower
maximum $X_{max}$, which is a logarithmically increasing function of the
primary energy. At fixed $E_0$, heavier primaries are expected to interact
earlier, thus giving a smaller value of $X_{max}$ (higher in atmosphere).
b) the muon and electron sizes of the showers. For a given primary energy,
EAS induced by heavy primaries develop earlier in atmosphere and less
energy is released
in the electromagnetic component, thus producing a smaller
$N_e$ at ground level, as compared to proton showers; on the other hand,
muons are produced more copiously in EAS by heavy primaries,
because of the higher number of low energy pions.\\
The EAS-TOP group studied the composition by analysing the behaviour of 
$\overline{N_{\mu}}$ as measured in vertical direction in narrow bins of
$N_e$, corresponding to $\Delta N_{e}/N_{e} = 12 \%$. The result is
shown in Fig.\ref{fi:etcomp}: data are compared with the results of a
full simulation including the detector response, where the 1 TeV
composition with equal slopes for all components was used, in this way
assigning constant composition  with energy.

\begin{figure}[htb]
 \begin{center}
 \mbox{\epsfig{figure=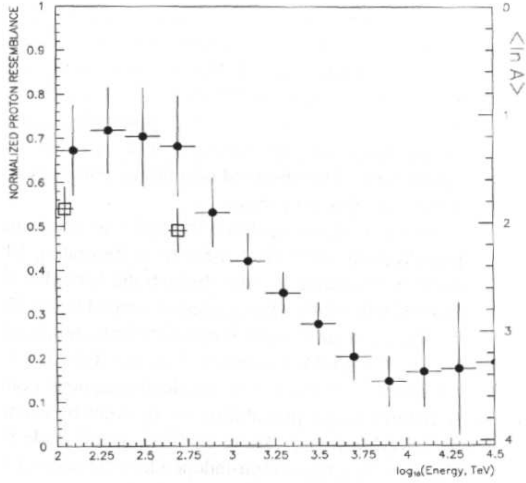,width=7cm,height=6.5cm}}
 \end{center} 
\vskip -1.0cm 
\caption{\em {Normalised proton resemblance plot from CASA-MIA. The open
squares give the estimated result if the composition is taken from JACEE
direct measurements \protect\cite{casa-comp}.}}
\label{fi:casares}
\end{figure}

The EAS-TOP data clearly suggest a growth of the mean A with energy, that is
a heavier composition above the knee. A change of $\Delta Log(N_{e})=0.5$
results in a $\Delta A/A \simeq 0.4$ \cite{et-comp}.\\
The ``K Nearest Neighbour'' test was used by CASA-MIA to study the
composition \cite{casa-comp}. 
Using the electron and muon densities at different distances from the core 
and the slope of the electron lateral distribution, samples of event for each
different primary mass are generated by Monte Carlo. An experimental event
is assigned to the ``light primary'' or ``heavy primary'' class by looking at 
the K nearest neighbours (KNN) in the plane of the used variables: the event 
will belong to the light primary group if e.g.more than $50 \%$ of its 
KNN are light primaries. Due to fluctuations, which tend to
superimpose classes, only broad classes of ``p-like'' and ``Fe-like''
events can be used. 
The proton resemblance, defined as the average fraction of K nearest
neighbours which are protons, is shown in Fig.\ref{fi:casares} for K=5,
normalised such that a pure proton composition would lay along
the top of the plot and a pure iron along the bottom border.

\begin{figure}[htb]
 \begin{center}
 \mbox{\epsfig{figure=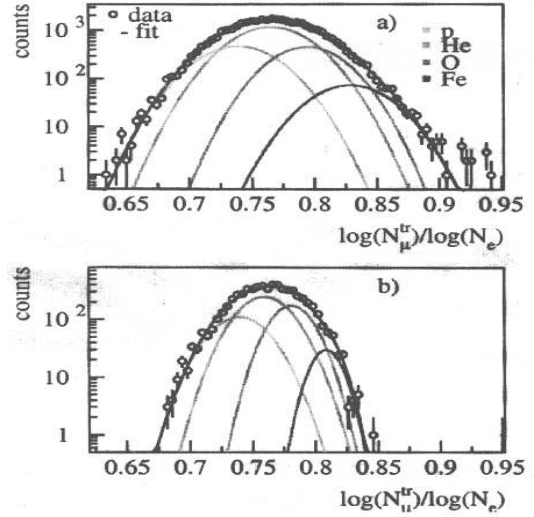,width=7.cm,height=7.0cm}}
 \end{center} 
\vskip -1.0cm
\caption{\em {$log \ N_{\mu}^{tr}/log \ N_{e}$ for two of the considered
energy bins: $6.2 \leq log(E/GeV) \leq 6.3$ and 
$6.7 \leq log(E/GeV) \leq 6.8$ \protect\cite{kasrap}.}}
\label{fi:kasrat}
\end{figure}

\begin{figure}[!htb]
 \begin{center}
 \mbox{\epsfig{figure=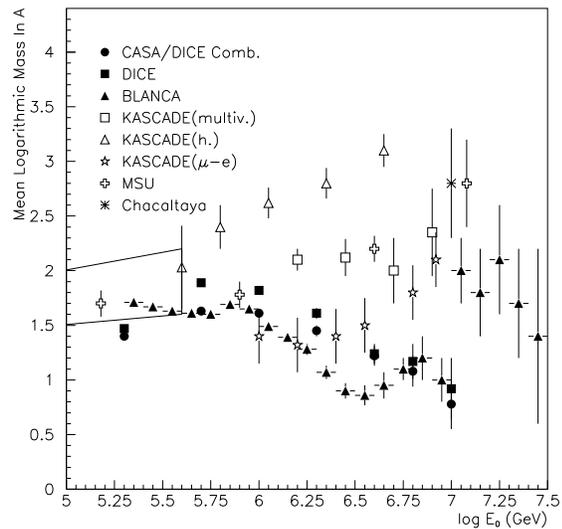,width=8cm,height=8cm}}
 \end{center} 
\vskip -1.0cm
\caption{\em {Mean logarithmic mass vs primary energy. The box represents
the region of direct measurements \protect\cite{wbso}. MSU data from \protect\cite{msu},
Chacaltaya data from \protect\cite{chaca}.}}
\label{fi:compall}
\end{figure}

The trend towards a heavier composition above the knee is evident; a
change in the hadronic interaction model used in the simulation does not
change the result. Classifying the events according to
their probability of being light or heavy primaries
and plotting the energy spectrum for the two classes separately, CASA-MIA
data suggest that the knee be due to the light mass group; the spectra are
consistent with the idea of cutoffs proportional to the particle rigidities.\\
The composition problem has been attacked by KASCADE people in a variety of
ways, using different observables and analysis methods. \\
The most sensitive dependence on primary mass was identified in the ratio
$log \ N_{\mu}^{tr}/log \ N_{e}$, which is found to be Gaussian
distributed at
fixed A \cite{kasrap}. The experimental ratio is fitted by a superposition
of simulated distributions (one for each primary mass group), directly
obtaining the fraction of each mass group, as shown in Fig.\ref{fi:kasrat}
for two energy bins. The
composition is dominated by the  light component up to about 4 PeV, getting
heavier above the knee; the analysis  also proves that the composition
cannot be described by a single component. The $<ln \ A>$ so obtained is shown
in Fig.\ref{fi:compall}.

\begin{figure}[!htb]
 \begin{center}
 \mbox{\epsfig{figure=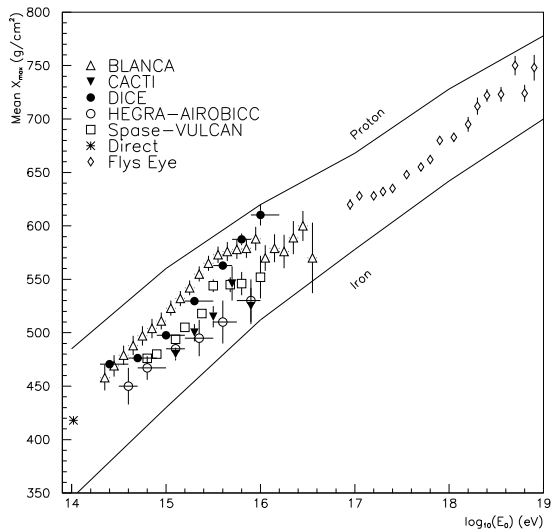,width=8cm,height=8cm}}
 \end{center} 
\vskip -1.0cm
\caption{\em {Mean height of shower maximum vs energy as measured by
various devices. HEGRA data from \protect\cite{heg_2}, Fly's Eye data from
\protect\cite{fly}. The lines show the expectations from a pure proton or iron
composition using CORSIKA+QGSJET \protect\cite{pryk}.}}
\label{fi:xmaxmed}
\end{figure}

A number of hadronic observables has also been used, such as the lateral
hadron distribution, the hadron energy spectrum, the maximum hadron energy
etc. in order to investigate the composition. As one can see in the world
survey given in Fig.\ref{fi:compall}, the hadronic data alone give a heavier
composition as compared to other data. \\
An interesting approach was used in \cite{kasrot}, where a multivariate
analysis using all the measured components of the EAS is performed. The result
shows a tendency to a lighter composition approaching the knee, followed by
an increase in the average mass above it.\\
A comparison among the results by KASCADE shows that the absolute scale
strongly depends on the observables which are used. It is clear that the
balance of energy among the different components of EAS in the simulation
does not reproduce the real situation; tests of the high energy interaction
models are been performed \cite{ka-had}.\\
In experiments like BLANCA, Spase-VULCAN \cite{spvul}, CACTI \cite{cacti},
Hegra-AIROBICC, $X_{max}$ is measured from the slope of the $\breve{C}$erenkov
lateral distribution, which is an almost linear function of
the depth of shower maximum. This function is rather independent on the
models chosen for the description of hadronic interactions, while any
interpretation of the experimental results in terms of primary composition
is not.\\
In the case of DICE, the imaging technique allows to measure $X_{max}$ in a
rather direct way, by fitting the shape of the shower $\breve{C}$erenkov
image in each of the 2 telescopes, knowing the arrival direction and the
core position of the shower. \\
A survey of the results is shown in Fig.\ref{fi:xmaxmed}, up to the Fly's
Eye energies (where air fluorescence is measured); the ``direct'' point
shows the $X_{max}$ that would be expected on the basis of balloon direct
measurements \cite{icrc93}. 

BLANCA data suggest a composition getting lighter
near the knee and turning to a heavier one after the knee energy. Data from
DICE require a composition becoming progressively lighter with increasing
energy. \\
DICE and CASA-MIA groups studied the composition problem also by means of a 
combination of measured parameters \cite{dice-gen}. Two estimates of the
mass have been derived, one using the $X_{max}$ as determined by 
DICE and the other with $N_{\mu}$ and $N_e$ by CASA-MIA, for each detected
shower. The combined use of different measurements allows first of all a
study of the systematics  biasing the composition results; moreover, the
requirement of consistency among various measurements allows to limit the
range of the parameters used in the models.
This analysis suggests a primary composition
becoming lighter at and above the knee, not excluding however a constant
composition around the knee energy.\\
A survey of the previously described results on primary composition in
terms of $<ln \ A>$ is shown in Fig.\ref{fi:compall}.

\section{Conclusion}

The cosmic ray energy spectrum and composition are studied with a
variety of experimental techniques detecting different air shower
components in the energy region above 1 TeV.\\
Below the knee, no new data are available and the conclusions reached by
JACEE and RUNJOB experiments still hold; however new projects, planned
to fly on balloons or on the Space Station, are in
progress. They will surely extend the explorable energy region and the
available statistics on single nuclei.\\
The energy spectrum has been studied in detail both by charged particles
and $\breve{C}$erenkov light ground arrays and some firm conclusions were
reached: all data
agree on the existence of the knee in the primary energy spectrum of cosmic
rays at an energy $\simeq 3-4 \ PeV$. The bend has been seen in all shower
components, thus supporting an astrophysical interpretation of the knee as
opposite to that of a change in the hadronic interaction picture at these
energies. All results agree to attribute the knee to the medium-light mass
primaries.\\
Progress has been made as regards the mass composition.
Almost all the ground array results show an increase in the primary mean
logarithmic mass above the knee, even if the absolute scale can be quite
different. 
There are however contradicting results coming from experiments relying on
the $\breve{C}$erenkov light detection from Air Showers; the differences in
the measure of $X_{max}$ are however quite big.\\
It is very important to study in detail the systematics which could bias
the results; the combined use of different observables and the comparison
among various data sets can help in this task.
The different sensitivity to composition of the various used observables,
the methods employed to determine the primary energy and the problems found
in the models used in the simulations could explain the spread in the results
which is apparent in Fig.\ref{fi:compall}. \\

{\bf Acknowledgements}

I would like to thank my friends G.Navarra, B.Alessandro and A.Chiavassa
for the interesting and useful discussions which were the basis of this
paper. Sincere thanks also to all my Brazilian friends, for their warm
hospitality during my stay in Campinas.

\end{document}